\documentclass[article,aps,onecolumn]{revtex4-2}
\usepackage{graphicx}
\usepackage{placeins}
\usepackage{float}
\usepackage{subfigure}
\usepackage{longtable}
\usepackage{amssymb}
\usepackage{filecontents}
\usepackage{natbib}
\usepackage[table]{xcolor}
\usepackage{setspace}
\begin{document}

\title{Structural properties of amorphous Na$_3$OCl electrolyte by first-principles and machine learning molecular dynamics}

\author{Tan-Lien Pham}
\altaffiliation{Current address: Department of Chemistry and Biochemistry Florida State University Tallahassee, FL 32304, USA}
\affiliation{Multiscale Materials Modeling Laboratory, Department of Physics, University of Ulsan, Ulsan 44610, Republic of Korea}
\author{Mohammed Guerboub}
\affiliation{Universit\'e de Strasbourg, CNRS, Institut de Physique et Chimie des Mat\'eriaux de Strasbourg, UMR 7504, F-67034 Strasbourg, France}
\affiliation{ADYNMAT CNRS consortium, F-67034 Strasbourg, France}
\author{Steve Dave Wansi Wendj}
\affiliation{Universit\'e de Strasbourg, CNRS, Institut de Physique et Chimie des Mat\'eriaux de Strasbourg, UMR 7504, F-67034 Strasbourg, France}
\affiliation{ADYNMAT CNRS consortium, F-67034 Strasbourg, France}
\author{Assil Bouzid}
\affiliation{Institut de Recherche sur les C\'eramiques, UMR 7315 CNRS-Univesit\'e de Limoges, Centre Europ\'een de la C\'eramique, 12 rue Atlantis 87068 Limoges Cedex, France}
\author{Christine Tugene}
\affiliation{Universit\'e de Strasbourg, CNRS, Institut de Physique et Chimie des Mat\'eriaux de Strasbourg, UMR 7504, F-67034 Strasbourg, France}
\author{Mauro Boero}
\altaffiliation{Current address: CNRS, Laboratoire ICube, Université de Strasbourg, UMR 7357, F-67037 Strasbourg, France}
\affiliation{Universit\'e de Strasbourg, CNRS, Institut de Physique et Chimie des Mat\'eriaux de Strasbourg, UMR 7504, F-67034 Strasbourg, France}
\affiliation{ADYNMAT CNRS consortium, F-67034 Strasbourg, France}
\author{Carlo Massobrio}
\altaffiliation{Current address: CNRS, Laboratoire ICube, Université de Strasbourg, UMR 7357, F-67037 Strasbourg, France}
\affiliation{Universit\'e de Strasbourg, CNRS, Institut de Physique et Chimie des Mat\'eriaux de Strasbourg, UMR 7504, F-67034 Strasbourg, France}
\affiliation{ADYNMAT CNRS consortium, F-67034 Strasbourg, France}
\author{Young-Han Shin}
\affiliation{Multiscale Materials Modeling Laboratory, Department of Physics, University of Ulsan, Ulsan 44610, Republic of Korea}
\author{Guido Ori}\email[]{Author to whom correspondence should be addressed: guido.ori@cnrs.fr}
\affiliation{Universit\'e de Strasbourg, CNRS, Institut de Physique et Chimie des Mat\'eriaux de Strasbourg, UMR 7504, F-67034 Strasbourg, France}
\affiliation{ADYNMAT CNRS consortium, F-67034 Strasbourg, France}

\begin{abstract}
Solid-state electrolytes mark a significant leap forward in the field of electrochemical energy storage, offering improved safety and efficiency compared to conventional liquid electrolytes. Among these, antiperovskite electrolytes, particularly those based on Li and Na, have emerged as promising candidates due to their superior ionic conductivity and straightforward synthesis processes. This study focuses on the amorphous phase of antiperovskite Na$_3$OCl, assessing its structural properties through a combination of first-principles molecular dynamics (FPMD) and machine learning interatomic potential (MLIP) simulations. Our comprehensive analysis spans models ranging from 135 to 3645 atoms, allowing for a detailed examination of X-ray and neutron structure factors, total and partial pair correlation functions, coordination numbers, and structural unit distributions. We demonstrate the minimal, albeit partially present, size effects on these structural features and validate the accuracy of the MLIP model in reproducing the intricate details of the amorphous Na$_3$OCl structure described at the FPMD level.
\end{abstract}

\maketitle

\section{Introduction}

Solid-state electrolytes are a focal point in the advancement of electrochemical energy storage technologies, offering the potential to address the safety and performance challenges inherent in traditional liquid electrolytes~\cite{zhao20, fam19}.
Notably, antiperovskite electrolytes, rich in Li and Na with low melting points, have been identified as a promising category due to their superior ionic conductivity and expedited synthesis processes~\cite{Dawson2021,Wei22,Shao2019,Kim2022,Yang2021}.
Antiperovskite electrolytes based on X$_3$OA (X= Li$^+$ or Na$^+$; A= halides (Cl$^-$, Br$^-$, I$^-$) or other anions (BH$_4^-$, NO$_2^-$)) have been synthesized in various forms, including purely crystalline, glassy, and hybrid amorphous/crystalline phases~\cite{Dawson2021,Dawson2022,Fan17,Den21,Wang2015,Oh2021,Dawson2018,Ahiavi2020,Xu2020,Zheng2021,Bra17}. In particular, glassy Li$_3$OCl- and Na$_3$OCl-based antiperovskites have shown remarkable conductivities of approximately $\sim$0.1 mS/cm at room temperature, with glass transition temperatures between 390 and 450 K~\cite{Bra14,Bra16a,Bra16b}. 
Also, amorphous Li$_3$OCl has been used as a matrix to embed Li-La-Zr-Ta-O garnet-type oxide particles, resulting in high room-temperature conductivity of 0.2 mS/cm and an extensive electrochemical stability window up to 10 V. The amorphous phase acted as a binder and filler, ensuring the formation of an integrated composite solid electrolyte with a continuous, widespread ionic conductive network. This significantly reduces the interfacial resistance with lithium metal anodes~\cite{Tia18,Tia2019,Gao2021,Tang22}.\\
However, despite such enormous potential for applications \cite{Dawson2021,Zheng2021,Han18,pham_2023,Ste18,Fri23}, a precise understanding of the structural and ion conduction mechanisms in these glassy antiperovskite electrolytes is not been achieved yet.
Indeed, recent literature on antiperovskites for solid-state batteries has highlighted the limited scope of structural characterizations available, underscoring the urgent need for a quantitative structural assessment to prevent misinterpretation of the relationship between structure and performance~\cite{Dawson2018}. While the structure and ion dynamics in crystalline compounds are well understood, amorphous antiperovskites await a clearcut structural characterization.
In this context, resorting to atomistic modeling has proved worthwhile to better understand the structural and transport properties of solid-state electrolytes~\cite{pol22}, being this approach complementary to experimental efforts.\\
A revealing example has been provided by our recent study of H-free sodium oxyhalide Na$_3$OCl and hydroxyhalide antiperovskites Na$_{3-x}$OH$_{x}$Cl ($x$~=~0.5, 1) using first-principles molecular dynamics (FPMD). Intricate details of their amorphous structure, the dynamics of Na ions, the role of H atoms, and their collective impact on ionic A revealing example has been provided by our recent study of H-free sodium oxyhalide Na$_3$OCl and hydroxyhalide antiperovskites using first-principles molecular dynamics (FPMD).  Intricate details of their amorphous structure, the dynamics of Na ions, the role of H atoms, and their collective impact on ionic conductivity have been highlighted~\cite{pham_2023b}. Due to intrinsic limits of our study based on a relatively small computational cell of  135 atoms, an extension to large sizes is appropriate to estimate  size-dependent effects  and obtain a more robust assessment of structural and dynamical properties, fully devoid of artifacts that might characterize the behavior of minimal size models. With this purpose in mind we focus here on amorphous Na$_3$OCl electrolyte by harnessing atomic-scale modeling through FPMD and classical MD simulations driven by a machine learning interatomic potential (MLIP). We considered models containing up to 3645 atoms fully representative of experimental behaviors.

To substantiate our structural analysis, we conduct an in-depth comparison with experimenta  X-ray and neutron structure by considering also the total pair correlation functions. While X-ray probing plays a crucial role, neutron studies are unavoidable to investigate quantitatively antiperovskites, especially due to their ability to detect low atomic number (Z) elements, which are beyond the capabilities of X-ray techniques using synchrotron sources~\cite{Gao2023,Shah21,Tur23}.
Furthermore, neutron diffraction and scattering studies are  gaining increased interest for oxyhalide materials, as they provide  insights into the arrangement of sodium and oxygen atoms, well beyond the level of detail achievable through X-ray methods~\cite{Gao2023,Shah21}. \\
The paper is structured as follows: Section II outlines the FPMD and MLIP modeling techniques and the systems studied, including the construction of the FPMD database and the evaluation of the developed MLIP against DFT data. Section III is devoted to the FPMD and MLIP results analysis, providing  the structural properties such as structure factors and total and partial pair correlation functions, coordination numbers, and structural units distributions. These findings are thoroughly analyzed and compared with existing experimental and modelling data for both Na$_3$OCl- and Li$_3$OCl-based systems. Concluding remarks are presented in Section IV.

\section{Computational methodology and models}
\subsection{First-principles molecular dynamic simulations}
The Car-Parrinello (CP) method~\cite{cp85} was used to produce trajectories of the targeted Na$_{3}$OCl
system. The exchange-correlation functional selected is the generalized gradient approximation (GGA) proposed
by Perdew, Burke and Ernzerhof (PBE)~\cite{pbe}.
The valence-core interaction was described by numerical norm-conserving Troullier-Martins (TM)~\cite{TM91}
pseudopotentials for all elements (Na, O, and Cl). In the case of Na, semi-core states were included to ensure a good description of the energetics and electronic features dependent on its cationic nature.
This amounts to electronic configurations \rm{[He]} 2s$^2$, 2p$^6$, 3s$^1$ for Na; \rm{[He] 2s$^2$, 2p$^4$ for O; and \rm{[Ne] 3s$^2$, 3p$^5$ for Cl. Valence electrons are treated explicitly and represented on a
plane-wave basis set 
with the sampling of the Brillouin restricted to the $\Gamma$ point.
A fictitious electron mass of 600 a.u. and a time step of 0.12 ~fs ensured optimal conservation of the constants of motion. 
FPMD simulations were performed in the canonical NVT ensemble with the ionic temperature controlled with a 
Nos{\'e}-Hoover~\cite{Nose1,Nose2,Nose3,Massobrio2022} thermostat chain.~\cite{Tucker92} 
For the simulations performed at high temperature ($>$1000 K) 
we used a Bl\"{o}chl-Parrinello electronic thermostat~\cite{Parr92} with a target kinetic energy of 0.1 a.u. to control the fictitious motion of the electronic degrees of freedom.
In our previous work~\cite{pham_2023b}, the Na$_3$OCl glass model was generated by quenching from the melt, starting from an initial configuration consisted of a Na$_3$OCl crystal cubic unit cell replicated $3\times 3\times 3$
times to obtain a 135 atoms model (81 Na, 27 O, and 27 Cl) in a cubic simulation cell of side 
13.6148 {\AA}~\cite{Hippler1990}. The initial box was expanded by 15\% with respect to crystal density in order to ease  melting at a high temperature.
The model underwent a thermal cycle via canonical NVT simulations, from T = 1200 K to T = 300 K along and finally equilibrated at T= 300 K for about  70 ps (see ref.~\cite{pham_2023b} for details). 
At the end of the cycle of T = 300 K, the density of the system was set to 2.04 g/cm$^3$ in order to reduce the residual stress to values very close to  0 GPa. Along the thermal cycle the energy  cutoff for the plane wave expansion was R$_{Ec}$= 80 Ry. 
In this study, we generated two novel amorphous models of Na$_3$OCl, consisting of 135 and 405 atoms, respectively, each possessing a density of 2.04 g/cm$^3$. Both models underwent a FPMD controlled thermal cycle following the following protocol: 3 ps at T = 300 K, 3 ps at T = 600 K, 3 ps at T = 900 K, 25.3 ps at T = 1200 K, 22 ps at T = 900 K, 22 ps at T = 600 K, and 30 ps at T = 300 K. The phase at T = 1200 K exhibited 
a diffusion coefficient of the order of 10$^{-5}$ cm$^2$/s, indicative of a complete loss of memory of the initial configuration.
Subsequent sections present the atomic structure results, reported as time-average values over the last 20 ps of the thermal cycles at T= 300 K. Specifically, the structural properties for the 135-atom FPMD size model (denoted as FPMD135 hereafter) represent the average of results from the new 135-atom model produced in this work and those from ref.~\cite{pham_2023b}. Similarly, the 405-atom model results correspond to the model produced in the present work, identified as the FPMD405 model.
For all the simulations performed here, we used the developer version of the CPMD package~\cite{cpmd}.\\

\subsection{DFT-FPMD database composition}
We build a database made of reference configurations on the desired regions of phase space and associated quantum mechanical data (energies, forces and virial stresses) by extracting 400 representative configurations from the FPMD trajectories obtained at different temperatures (1200, 800, 450 and 300 K) for the 135-atoms model produced in ref.~\cite{pham_2023b}. This amounts to a total of 400 total energy values, 162000 force components and 3600 virial components. In order to achieve a good accuracy of the database, we recomputed DFT energies, forces, and virials for all the configurations at an energy cutoff of 160 Ry. The built database is then split in 50:50 for training and testing sets.

\subsection{Machine learning interatomic potential: model fitting and training database}
We employ a kernel-based MLIP method using the Gaussian Approximation Potential (GAP) scheme, enabling the learning and reproduction of smooth, high-dimensional potential energy surfaces through the interpolation of DFT data~\cite{Bar10,Bernstein2019,Bar14,Bar15}. This MLIP scheme has demonstrated success in various applications, encompassing liquids, crystalline phases, as well as mono and binary glasses~\cite{del20,unruh22}.
However, its application to ternary amorphous systems has been relatively limited so far.
In the subsequent section, we briefly outline key aspects of the GAP methodology. Additional details regarding the GAP schemes can be found in dedicated reviews~\cite{Bar10,Bar15,Der2021,kla23}.
The GAP scheme used in the present work is based on the system's total energy expressed as the combination of a repulsive two-body potential and a many-body kernel based on the Smooth Overlap of Atomic Positions (SOAP) descriptor~\cite{Bar13}.  
\begin{equation}
E\equiv  \sum\limits_{i<j}^{n} \bold{V}^{(2)}(r_{ij})+ \sum\limits_{i}^{n} {\epsilon (\bold{q}_{i})},
\end{equation}
where, $i$ and $j$ index the atoms in the atoms, $V^{(2)}$ is the 2-body repulsive pair potential, $r_{ij}$ is the distance between atoms $i$ and $j$, and $\epsilon$($\bold{q}_{i}$) is the energy contribution of the $i^{th}$ atom and $\bold{q}_{i}$ is a descriptor characterizing the local atomic environment around the $i^{th}$ atom. The local atomic energy contribution $\epsilon$($\bold{q}_{i}$) is given by a linear combination of kernel functions,
\begin{equation}
{\epsilon (\bold{q}_{i})} =  \sum\limits_{j} \alpha_jK(\bold{q}_{i},\bold{q}_{j}),
\end{equation}
where, $K$ is a fixed kernel function quantifying the degree of similarity between atomic environment described by $q_i$ and $q_j$. 
The coefficients $\alpha_j$ are determined by a regularised linear regression of the energies, forces
and virials of the system computed with Eq.(2) that
are parametrized by these $\alpha_j$ values, and compared to
the corresponding energies, forces and stresses computed by DFT for the same structures. In SOAP, the local environment of the  $i^{th}$ atom is represented by its neighbour density function within a given cutoff $r_{cut}$,
\begin{equation}
\rho_i(\bold{r}) =  \sum\limits_{j} f_{cut}(r_{ij})exp
 \left[  -\frac{ (\bold{r}_i - \bold{r_{ij}})^2 } { 2\rho^2_{at} }  \right],
\end{equation}
where, $\bold{r}_i$ denotes the position vector of atom $i$ and $\rho_{at}$ controls the smoothness of the potential. Here, $f_{cut}$ is a cutoff function going smoothly to 0 at $r_{cut}$~\cite{liu20}.
Atomic neighbor densities are then expanded in a local basis set of orthogonal radial basis functions $g_n$(r) and spherical harmonics $Y_{lm}$ as:
\begin{equation}
\rho_i(\bold{r}) =  \sum\limits_{n<n_{max}}\sum\limits_{l<l_{max}} \sum\limits_{m=-l}c^i_{nlm}g_n(r)Y_{lm}\left( \bold{\hat{r}}\right),
\end{equation}
where $c^i_{nlm}$ are the expansion coefficients. The spherical power spectrum of these expansion coefficients then forms the descriptor as:
\begin{equation}
(\bold{q}_i)_{nn'm} =  \frac{1}{\sqrt{2l+1}}
\sum\limits_{m}(c^i_{nlm})^*c^i_{n'lm},
\end{equation}
which is  translational, rotational, and permutational invariance. For the many-body SOAP descriptor, the dot product kernel is determined by
\begin{equation}
{K (\bold{q}_i,\bold{q}_j)} =  
 \left|  -\frac{ (\bold{q}_i \cdot \bold{q_{j}}) } { |\bold{q_i}| \cdot | \bold{q_j}|}  \right|,
\end{equation}
The positive integer $\zeta$ allows to improve the sensitivity of the kernel and also increase the body order of the model. To reduce the computational cost, a sparsification method is used for the SOAP kernel.  
Overall, the SOAP descriptor represents atomic geometries by using a localized expansion of a Gaussian-smeared atomic density. In so doing, the estimation of a specific physical property is broken down into individual contributions
centered on atoms. These contributions effectively capture correlations between atoms inside each
localized environment. The assumption of locality, justified by the nearsightedness principle~\cite{prodan2005}
of electronic matter, is very advantageous as it effectively reduces the complexity of the regression
task, often encountered in machine learning schemes. The SOAP descriptor facilitates a thorough characterization of short- to medium-range order, allowing the identification of features that assess varying degrees of (dis)order in a given material~\cite{Bernstein2019,pham_2023b}.\\
The hyperparameters used in this work for constructing the descriptors, the kernel functions, and training a GAP are summarized in Table S1 of the Supplementary Material. 

\begin{figure*}[]
\includegraphics[width=18cm]{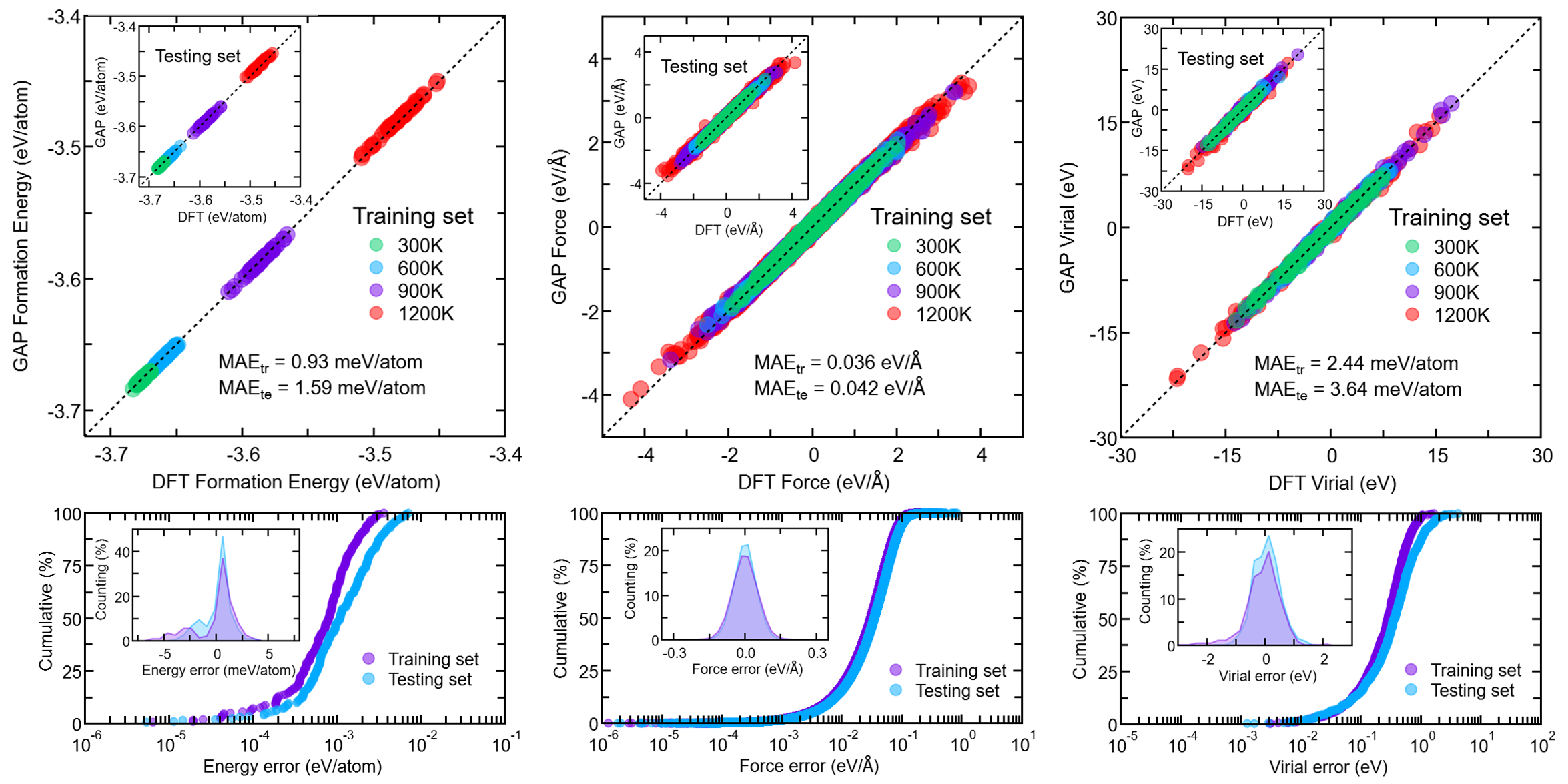}
\setstretch{2} 
\caption{Top: Scatter plots of DFT-computed and GAP-predicted total energies (left), forces (centre) and virials (right) for the training and testing (insets) sets of 200 configurations each. Bottom: Cumulative error distributions: a given point ($x$,$y$) on the curve indicates that $y$ percent of all structures have an error equal to or below $x$. Insets: Absolute errors distributions of the respective quantities obtained for the training and testing sets.}
\label{fig:MLvsDFT}
\end{figure*}

\subsection{MLP-GAP model performance assessment: errors for testing versus training sets}
We here assess the validation of the accuracy of our GAP model based on
the DFT-FPMD reference data. The training and testing datasets contain the energy, forces and virial components of 200 configurations each. As shown in Fig. 1, the total energies, atomic forces and virial as predicted by our model are compared with those from DFT-FPMD. Our GAP model successfully reproduces the energies with a low mean average error (MAE) of 1.59 eV/atom for testing datasets, significantly below the commonly referenced threshold of 5 meV/atom, which is often cited as indicative of a high-performing MLIP~\cite{siva20,der18,bar18}. Forces in the datasets are predicted with a
MAE of 0.04 eV/Å  whereas virials are predicted with a MAE of 3.64 eV, respectively. The results demonstrate that our GAP model is a good representative of the first-principles potential energy surface for amorphous Na$_3$OCl.
Figure 1 displays errors through counting and cumulative distribution plots. The counting distribution reveals a very tight overlap, while the cumulative distribution shows curves closely aligned for forces  and slightly shift to the right and downward (indicating slight increased confidence) when comparing predictions for energy and virials from training and testing datasets. Overall, across all three observables, we note an exceptionally close match, highlighting the smooth performance and first-principles accuracy of the developed MLIP-GAP.

\subsection{MLP-GAP molecular dynamics simulations}
We exploited the fitted GAP potential to produce new models of amorphous Na$_3$OCl by following a thermal cycle via canonical NVT simulations, according to the following protocol: heating up from T = 300 K to T = 1200 K along 5ps, melting at T = 1200 K for 100ps, cooling from T= 1200 K to T = 300 K with a cooling rate of 4 K/ps, and final equilibration at T = 300 K for 200ps.
Subsequent sections present the  structural analysis, reported as time-average values over the last 100 ps of the thermal cycles at T= 300 K. The equations of motion were integrated by using a timestep of 0.5-1 fs and NH thermostat was used as implemented in the LAMMPS code~\cite{lammps22}.  With such procedure we produced MLIP-GAP models with the same density (2.04 g/cm$^3$) for following sizes: 135 atoms model (hereafter denoted as GAP135 model), 405 (GAP405), 1080 (GAP1080) and 3645 (GAP3645) atoms models. GAP135 was averaged over 11 replica run whereas the other GAP models were averaged over 2 run. Figure 2 shows snapshots and details of the models simulated by FPMD and GAP.
The GAP models obtained have been employed to reinforce the structural findings of amorphous Na$_3$OCl derived from FPMD. This is achieved by leveraging large-scale models refined to DFT accuracy, thereby highlighting any potential size-related effects encountered. Wherever pertinent, the outcomes of the GAP models are thoroughly examined and discussed within the paper.

\begin{figure}[]
\centering\includegraphics[width=0.7\linewidth]{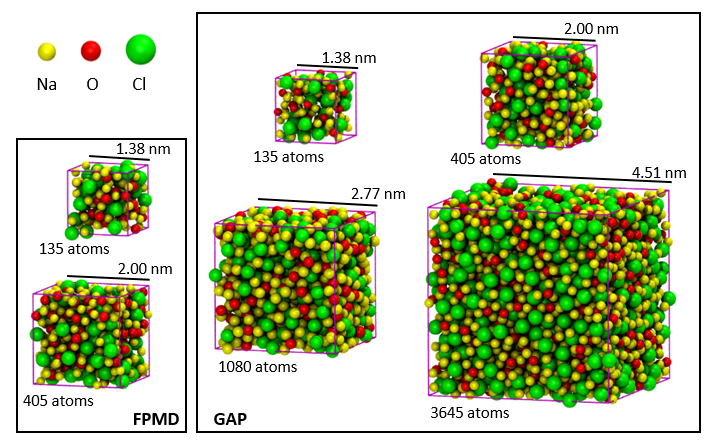}
\setstretch{2}
\caption{(Color online) Supercell amorphous models of Na$_3$OCl simulated by FPMD (135 and 405 atoms) and GAP (135, 405, 1080 and 3645 atoms). Color legend: Na, yellow; O, red; Cl, green.}
\label{fig:models}
\end{figure}

\begin{figure*}[t]
\begin{center}
\centering\includegraphics[width=1\linewidth]{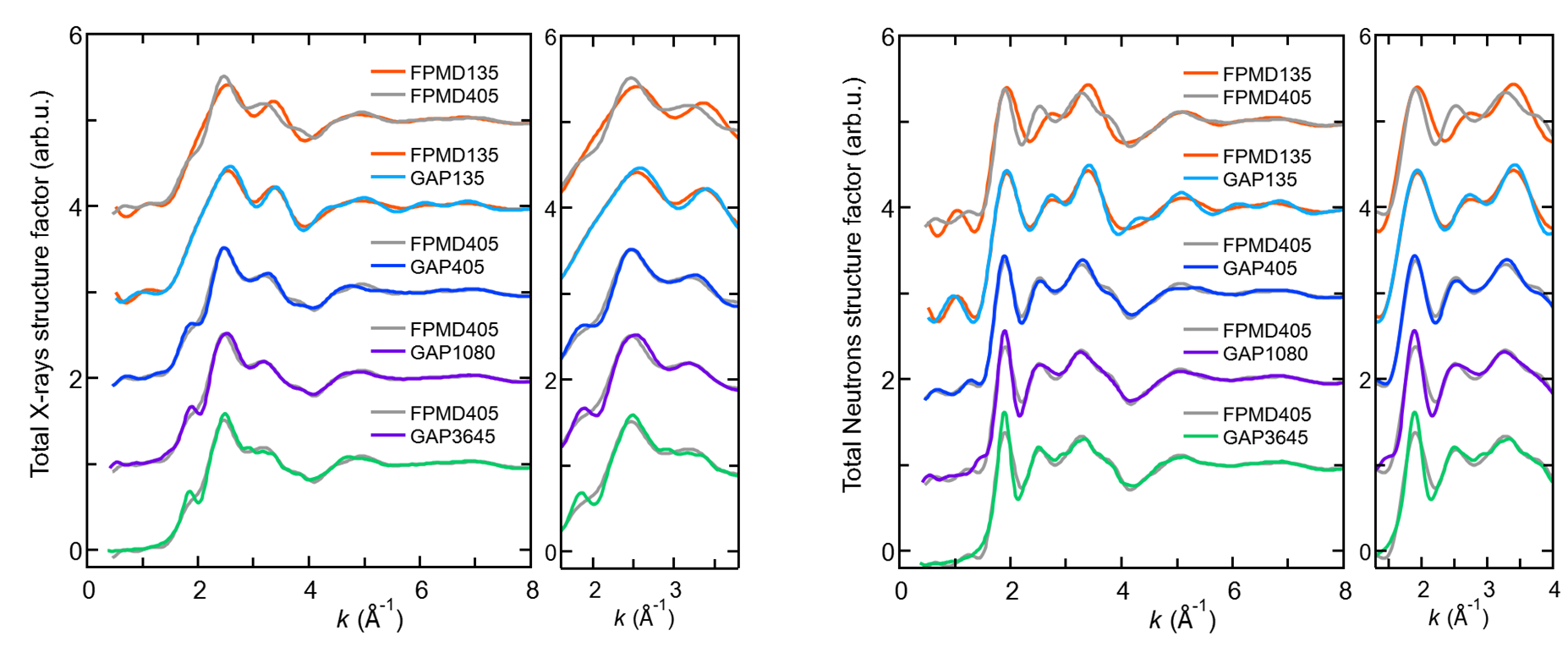}
\setstretch{2}
\caption{Total X-ray (right) and neutron (left) structure factors, \( S_T^x(k) \) and \( S_T^n(k) \), for amorphous Na\(_3\)OCl at 300K, are displayed for varying model sizes. FPMD data are depicted in orange and grey for the 135 and 405 atoms model, respectively. GAP data are indicated by light blue, blue, purple, and green  for the 135, 405, 1080, and 3645 atoms model, respectively. For \( S_T^x(k) \) and \( S_T^n(k) \), zoom-in in the 1.5-3.8~\AA$^{-1}$ and 1.4-4~\AA$^{-1}$ ranges are reported, respectively.
}
\label{FIG:sk}
\end{center}
\end{figure*}

\section{Results and Discussion}
In the following sections, we report a detailed analysis of various structural properties using the results obtained from both FPMD and MLIP schemes. Each analysis begins with an evaluation of size effects at the FPMD level by comparing models containing 135 and 405 atoms. Next, we compare the performance of MLIP against FPMD findings by assessing the congruence between the structures of the 135 and 405 atom models. Lastly, we extend the size effect assessment to the MLIP level, incorporating models with up to 3645 atoms.

\subsection{Total X-ray and Neutron structure factors}

The calculated total X-ray and neutron structure factors, denoted as \( S_T^x(k) \) and \( S_T^n(k) \), for amorphous Na$_3$OCl models are depicted in Fig. \ref{FIG:sk}. These were obtained through the Fourier transform (FT) of the real-space pair distribution functions generated by the simulations. While they exhibit broadly similar profiles, notable differences are observed between the structure factors of the 405-atom models and those of the 135-atom models, particularly regarding the peaks' maximum intensity and positions. For $S_T^x$(k), the primary peak at 2.5~\AA$^{-1}$ is more intense and goes along with a more pronounced shoulder at 1.9~\AA$^{-1}$ in the 405-atom models. Additionally, the secondary peak, located at 3.2~\AA$^{-1}$ in the 135-atom models, shifts to a slightly lower $k$-value of 3.1~\AA$^{-1}$ and exhibits a less defined minimum between the first and second peaks in the larger models. Following the initial peaks, both the 135-atom and 405-atom models exhibit much broader maxima and minima and rapidly damped
oscillations up to $k\sim$8~\AA$^{-1}$. These features of the structure factor curves are consistently observed across both model sizes.
In terms of \( S_T^n(k) \), the differences between the 135-atom and 405-atom models are more pronounced compared to those observed for \( S_T^x(k) \). Particularly, the 405-atom model shows the absence of the small peak at \( k \sim 1~\text{\AA}^{-1} \), indicating a vanishing intermediate-range order in this amorphous system. Concerning the peaks found at \( k < 4~\text{\AA}^{-1} \) in the 135 atoms model, the 405-atom model features a first peak with very similar maximum intensity and position at \( \sim 2.0~\text{\AA}^{-1} \), whereas the second peak appears at a lower \( k \) value (\( \sim 2.8~\text{\AA}^{-1} \) compared to \( \sim 3.0~\text{\AA}^{-1} \)) and is more pronounced. The third peak is also shifted to a lower \( k \) (\( \sim 3.4~\text{\AA}^{-1} \) versus \( \sim 3.6~\text{\AA}^{-1} \)) and is slightly less intense. Additionally, a distinct shoulder at \( \sim 3.8~\text{\AA}^{-1} \) is observed in the 405 atoms model, which is not present in the 135 atoms model. The 405-atom FPMD model, currently the largest FPMD model developed to date, represents our most accurate description of amorphous Na$_3$OCl. It aligns closely with the reported data from the 135-atom model, both from this study and previously reported~\cite{pham_2023b}, albeit with minor yet notable differences in peak positions and intensities, denoting minimal size effects within the size range of 135-atom to 405-atom models.\\
To further assess the degree of agreement between the FPMD models of 135 and 405 atoms, we employ the \(R_{\chi}\) factor~\cite{wright1993}. This factor is calculated using the equation defined in (\ref{eqt:Rx}), allowing for a quantitative analysis of the similarity between the calculated \(S_T^x(k)\) and \(S_T^n(k)\) of the two models and the impact of system size on structural features.
\begin{equation}
R_{\chi} = \left( \frac{\sum_{i}~ \left[S^{\rm fpmd}(k_i) - S^{\rm fpmd'}(k_i)\right]^2 }{\sum_{i} ~\left[S^{\rm fpmd}(k_i)\right]^2}\right)^{1/2}
    \label{eqt:Rx}
\end{equation}
In this formula, $S^{\rm fpmd}(k_i)$ and $S^{\rm fpmd'}(k_i)$ refer to the calculated FPMD structure factor of the 405 
 and 135 atoms models, respectively, at a given wave vector $k_i$.
Table~\ref{tableRx} details the calculated \(R_{\chi}\) values for both \(S_T^x(k)\) and \(S_T^n(k)\). Notably, within glasses simulated using FPMD, \(R_{\chi}\) values equal of 10 or lower in the 1-10 \(k\) range are commonly accepted as denoting quantitative agreement with experimental data~\cite{bouzid2021}. The computed \(R_{\chi}\) values for the 135-atom FPMD models, being 4.7 for X-ray and 7.5 for neutron structure factors, corroborate the above analyses, showing a reasonably good agreement with the data from the 405-atom model. However, these values also indicate a slightly greater discrepancy for \(S_T^n(k)\) compared to \(S_T^x(k)\), underscoring the nuanced impact of structure factor type on model agreement.\\
When comparing the total structure factors \( S_T^x(k) \) and \( S_T^n(k) \) for the 135-atom model using GAP and FPMD schemes, remarkably similar outcomes are observed. 
The consistency in data between FPMD and GAP is also equally evident in the 405-atom model for both \(S_T^x(k)\) and \(S_T^n(k)\) total structure factors. 
This similarity highlights the effectiveness of the developed MLIP potential in accurately capturing the structural features of amorphous Na\(_3\)OCl, closely mirroring the predictions of the more computationally accurate and costly FPMD approach. The high degree of quantitative agreement between GAP and FPMD schemes is further underscored by the low \(R_{\chi}\) values (Table~\ref{tableRx}). These values, which are below 3 for X-ray and below 4 for neutron structure factors in the 135-atom models, and less than 3 for both X-ray and neutron in the 405-atom models, highlight the FPMD (DFT)-accuracy of the GAP model in replicating the detailed structural features of amorphous Na\(_3\)OCl.\\
Having established the FPMD (DFT) accuracy of our newly developed MLIP GAP potential, we are in a position to explore size effects by analyzing the 1080 and 3645-atom models obtained with GAP. By comparing these profiles to the 405-atom model via FPMD and GAP, we observe only very minor differences. For \(S_T^x(k)\), the subtle shoulder at 1.9 Å\(^{-1}\) previously noted in the 405-atom model becomes a discernible, albeit small, peak with increased intensity in the 1080 and 3645-atom models. Similarly, the intensity of the first peak maximum in \(S_T^n(k)\) increases in the models with 1080 and 3645 atoms compared to the 405-atom model, highlighting subtle but notable size-dependent effects.\\

\begin{figure*}[t]
\begin{center}
\centering\includegraphics[width=0.9\linewidth]{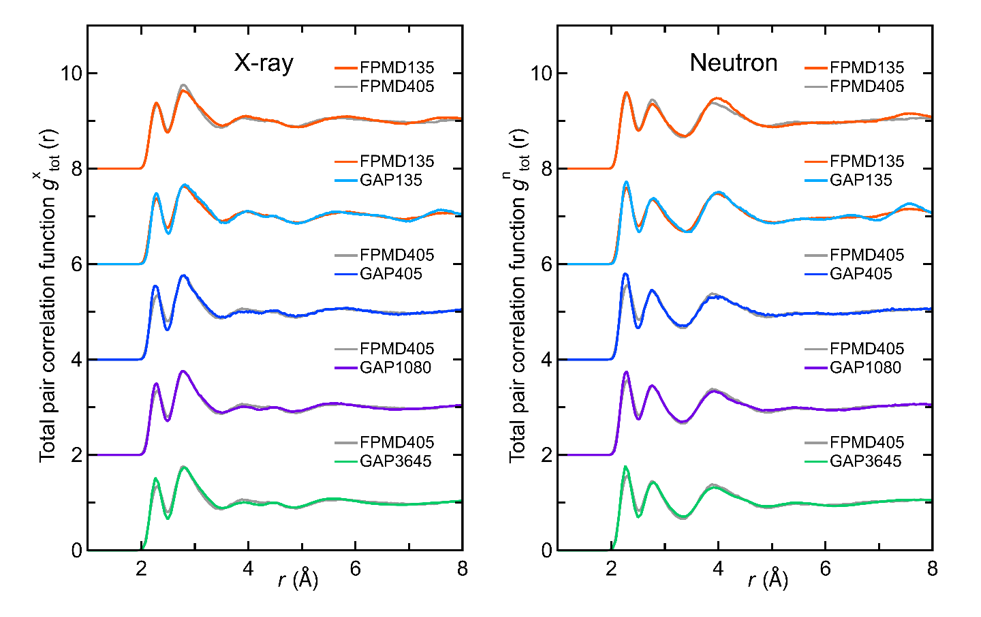}
\setstretch{2}
\caption{
Calculated X-ray (right) and neutron (left) total pair correlation functions, \( g_{t}^{x}(r) \) and \( g_{t}^{n}(r) \), for amorphous Na\(_3\)OCl models at 300K. FPMD data is reported for 135 (orange) and 405 (grey) atoms models, whereas GAP data is reported for 135 (light blue), 405 (blue), 1080 (purple), and 3645 (green) atoms models.
}
\label{gr_tot}
\end{center}
\end{figure*}

\begin{table}[]
\setstretch{2}
\caption{\small Averaged goodness-of-fit \(R_{\chi}\) parameters for the 135 atoms FPMD model and GAP models ranging from 135 to 3645 atoms, indicating their degree of agreement with the reference 405 atoms FPMD model data in terms of neutron and X-ray structure factors (\(R_n^{S(k)}\) and \(R_x^{S(k)}\), respectively) and the total pair correlation functions (\(R_n^{T(r)}\) and \(R_x^{T(r)}\), respectively). \(R_n,x^{S(k)}\) and \(R_n,x^{T(r)}\) are calculated over the ranges of 0.75-10.0 Å\(^{-1}\) and 0.75-8.0 Å, respectively. For the 135 atoms GAP model, the \(R_{\chi}\) values are provided with reference to the 135 atoms FPMD model data, serving for assessing the consistency of the GAP potential across different model sizes.}
\begin{tabular}{l|c|cccc}
\hline
\hline
	 &  FPMD & \multicolumn{4}{c}{GAP} \\
	n. atoms & 135  & 135& 405 & 1080 & 3645\\
\hline
X-Rays  &  &     & &&\\
$R^{\rm x}_{\rm S(k)}$   &  4.7    &2.7 &2.6&3.5&3.6\\
$R^{\rm x}_{\rm T(r)}$    &  4.9   & 4.8 &5.4& 3.6&4.3\\
\hline
Neutrons  &  &   & &&\\
$R^{\rm n}_{\rm S(k)}$    &  7.5   &3.9&2.6&5.4&5.3\\
$R^{\rm n}_{\rm T(r)}$  &  5.8    &5.7&5.3&3.9&4.3\\

\hline
\hline
\end{tabular}
\label{tableRx}
\end{table}
\setlength{\belowcaptionskip}{0pt}

\subsection{Total and partial pair-correlation functions}
The X-ray and neutron total pair correlation functions, \( g_{\text{tot}}^{x/n}(r) \), for Na\(_{3}\)OCl models obtained via FPMD and GAP, are shown in Fig.~\ref{gr_tot}. These functions are characterized by two primary peaks at about 2.28 Å and 2.77 Å, with distinct intensities between X-ray and neutron data. Furthermore, \( g_{\text{tot}}^{n}(r) \) exhibits an additional, broader peak around 4 Å, followzed by reduced oscillations at larger distances. The presence of distinct peaks in the total pair correlation functions at specific interatomic distances indicates a consistent short-range order within the amorphous Na$_3$OCl structure~\cite{pham_2023b}. The additional broader peak observed in neutron correlation functions implies differences in the way X-ray and neutron scatterings are sensitive to the specific atomic arrangements.The damped oscillations in the total pair correlation functions, \( g_{\text{tot}}^{x/n}(r) \), for distances greater than 5 Å lend support to the previous observations derived from the structure factors \( S_T(k) \). This behavior is indicative of a very limited intermediate range order within the amorphous Na\(_{3}\)OCl structure, which aligns with the characteristics typically associated with amorphous antiperovskites~\cite{pham_2023b}.
The close agreement between the 135 and 405-atom FPMD models implies a negligible size effect on these correlations at the FPMD level. This is corroborated by the relatively low \( R_{\chi} \) values reported in Table~\ref{tableRx}, with a \( R^x_{T(r)} \) of 4.9 and a \( R^n_{T(r)} \) of 5.8 for the FPMD 135-atom model when compared to the 405-atom model. The remarkable quantitative agreement between the GAP and FPMD results is assessed by \( R_{\chi} \) values spanning 4.8 to 5.4 for X-rays and approximately 5.7 to 5.3 for neutrons, for the 135 and 405-atom models, respectively.
The GAP models with 405, 1080, and 3645 atoms are characterized by highly similar \( g_{\text{tot}}^{x/n}(r) \) for both X-ray and neutron, suggesting a minimal size effect on these correlation functions. However, when considering the \( R_{\chi} \) values, the concordance with FPMD data improves for the larger models, as shown by the reduced \( R_{\chi} \) values, which are close to $\sim$ 4. This reflects an enhanced accuracy in structural reproduction as the GAP model size increases.\\
Analyzing the partial pair correlation functions g$_{\rm{\alpha \beta}}$(r) (Fig.~\ref{gr_p}) allows us to trace back the role of each individual pair contribution with respect to the total one. In Tab.~\ref{dist} we report the nearest-neighbor distances $r_{ij}$ identified by the position of the first maximum of the g$_{\rm{\alpha \beta}}$(r) for the Na$_{3}$OCl models simulated by FPMD and GAP. For comparison, we also report the nearest-neighbor distances found in synthesized pure crystalline Na$_3$OCl~\cite{Pham2018,pham_2023}.
\begin{figure*}[t]
\begin{center}
\centering\includegraphics[width=1\linewidth]{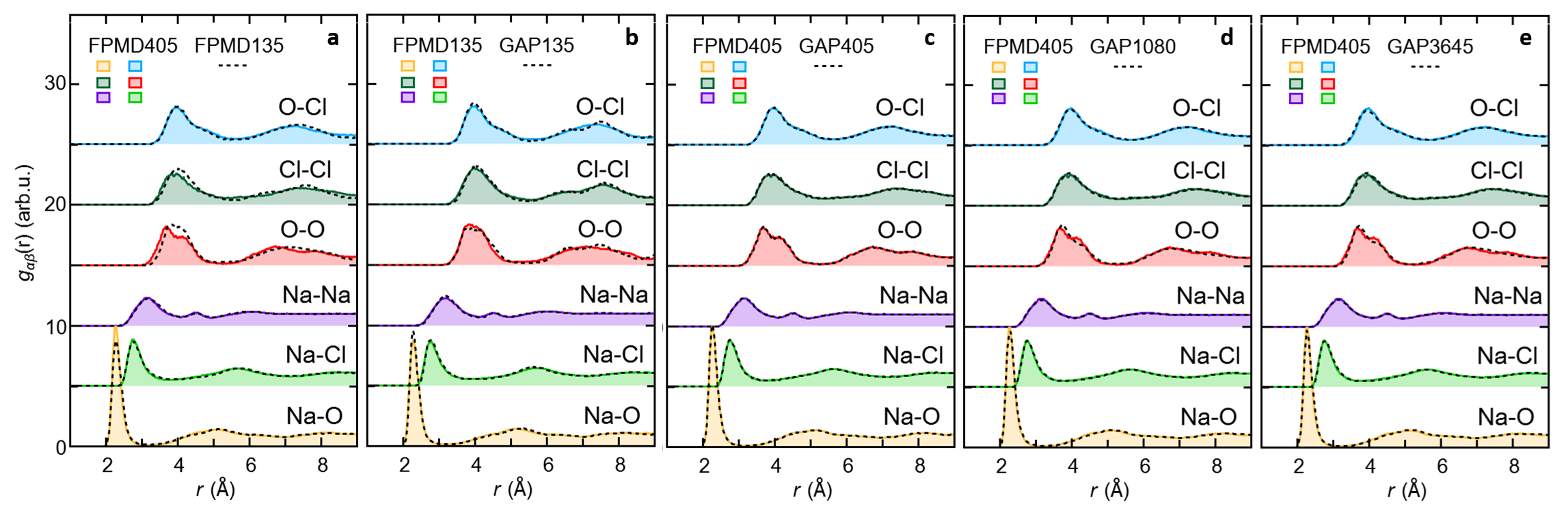}
\caption{The partial pair correlation functions \( g_{\text{NaO}}(r) \), \( g_{\text{NaCl}}(r) \), \( g_{\text{NaNa}}(r) \), \( g_{\text{OO}}(r) \), \( g_{\text{OCl}}(r) \), and \( g_{\text{ClCl}}(r) \) for amorphous Na\(_3\)OCl at \( T = 300 \) K are shown, as obtained comparing 135 and 405 atoms FPMD models (a), 135 atoms FPMD and GAP models (b), and 405 atoms FPMD model with 405 (c), 1080 (d) and 3645 (d) atoms GAP model. The plots are vertically offset for better visibility.
}
\label{gr_p}
\end{center}
\end{figure*}
\begin{table*}[]
\setstretch{2}
\caption{\small
Upper part: nearest-neighbour interatomic distances $r_{ij}$ (in {\AA}) identified by the position of the first maximum of the pair correlation functions $g_{ij}(r)$. For comparison, we report the values of the crystalline Na$_3$OCl phase obtained by experiments.\cite{Hippler1990,Pham2018} The average statistical error for the simulated $r_{ij}$ data (FPMD and GAP), is 0.02~{\AA}. Center part: total and pair coordination numbers as defined by taking different cutoff radii for the definition of a total or a ``cation-anion'' shell of interactions. Values in parenthesis correspond to the statistical uncertainty on the last reported digits.
Lower part: density ($d$, g/cm$^3$) of the models simulated. For GAP models $d$ is computed along 100~ps at 300~K and 1~Katm performed in the NPT ensemble.
 GAP models data is averaged over 11 replica for 135-atom model and over 2 replica for the other GAP models.}
\begin{tabular}{l|c|cc|cccc}
\hline
\hline
	&  \multicolumn{1}{c}{Exp.\cite{Hippler1990,Pham2018}} &  \multicolumn{2}{c}{FPMD} & \multicolumn{4}{c}{GAP} \\
 	 & Cr.     &Am.  & Am. & Am. & Am. & Am. & Am.\\
	n. atoms & --                            &135  & 405 & 135 & 405 & 1080 & 3645\\
\hline
$r_{\rm NaO}$    & 2.25   &    2.28&2.30    & 2.28&2.27&2.27&2.26\\
$r_{\rm NaCl}$   & 3.18   &   2.77 &2.75   & 2.75&2.76&2.78&2.79\\
$r_{\rm NaNa}$   &  3.18    &    3.19& 3.11 &3.16& 3.15&3.13&3.14 \\
$r_{\rm OO}$     &  4.50  &   3.83 & 3.79   & 3.81&3.72&3.74&3.73\\
$r_{\rm OCl}$    &  3.89   &    3.90&3.94   & 3.95& 3.95&3.95&3.93\\
$r_{\rm ClCl}$   &  4.50   &  3.96  &3.89   & 4.00& 3.86&3.93&3.94 \\
\hline
$n^t_{\rm Na}$   & 14   &   8.06(4) & 8.16(3)   & 8.16(9)& 8.23(2)& 8.24(3)&8.24(2)\\
$n^p_{\rm Na}$    & 6   &   4.0(1)  & 4.10(2)   & 4.06(5)& 4.03(3)& 4.09(2) & 4.09(1)\\
$n^t_{\rm O}$   &  6    &   6.40(3) & 6.53(3)   & 6.34(8)& 6.44(3)& 6.46 (3)&6.40(2)\\
$n^t_{\rm Cl}$   & 12   &   5.75(4) & 5.84(4)   & 5.85(7)& 5.79(3)& 5.88 (3) &5.87(2)\\
\hline
$d$ (g/cm$^3$)     & 2.20 & 2.04 &2.04 & 2.01 & 2.02  &2.01&2.02\\
Error(\%) & -- &-- & -- &  (-1.3\%) &(-1.0\%)& (-1.3\%) &(-1.0\%)\\
\hline
\hline
\end{tabular}
\label{dist}
\end{table*}
\setlength{\belowcaptionskip}{0pt}
The first peak of \( g_{\text{tot}}(r) \) for amorphous Na\(_3\)OCl FPMD models can be attributed entirely to the Na-O pair at approximately 2.28-2.30 Å, while the second peak is primarily due to the Na-Cl pair at roughly 2.75-2.77 Å, with only partial contributions from Na-O and Na-Na pairs within 3.11-3.19 Å in the \( g_{\alpha \beta}(r) \) functions. The amorphous nature of the network favors relatively short Na-Cl distances ($\sim$2.75-2.77 \AA) when compared to those found in the crystalline phase, where each Cl atom is surrounded by twelve Na atoms at a distance of $\sim$3.18 \AA. 
The initial peak of \( g_{\text{NaNa}}(r) \) is responsible for the broadened profile of the second peak observed at about 2.8-3.1 Å in the total pair correlation functions. Upon analyzing the peak positions in \( g_{\text{NaO}}(r) \) for amorphous Na\(_3\)OCl, the Na-O bond distance is slightly larger than the typical ionic bond seen in crystalline Na\(_3\)OCl, which is around 2.25 Å, whereas the Na-Cl ionic bond distance is shorter compared to the crystalline counterpart at approximately 3.18 Å. The anion-anion pair correlation functions, \( g_{\text{OO}}(r) \), \( g_{\text{ClCl}}(r) \), and \( g_{\text{OCl}}(r) \), exhibit a principal peak centered near 3.8-4.0 Å. Both the 135 and 405 atoms FPMD models shows very similar partial pair correlation functions \( g_{\alpha \beta}(r) \), with the 405 atoms model having a slightly narrower first peak for the Na-O pair, a slightly shorter Cl-Cl (3.9 Å vs 4.0 Å) and more distinct first peaks for O-O. In the 135 atoms model, the O-O pair presents a broadened peak centered around 3.83 Å, while in the 405 atoms model, it displays a primary peak at approximately 3.79 Å with an additional shoulder at around 4.1 Å. Notably, the O-O pair interaction is of crucial significance, as the interaction between O atoms has been identified as critical for the electronic and electrochemical stability of amorphous Li\(_3\)OCl in contact with a Li metal electrode~\cite{Choi2022}.\\
The GAP models accurately reproduce the \( g_{\alpha \beta}(r) \) functions observed in both the 135 and 405 atoms FPMD models, further validating the FPMD (DFT)-level accuracy of the GAP approach developed for amorphous Na\(_3\)OCl. Notably, the discrepancies between the 135 and 405 atoms FPMD models are precisely captured by the GAP models, confirming that these variations can be ascribed to size effects. The GAP calculations for models containing 405, 1080, and 3645 atoms yield remarkably consistent results across all pairs, with the exception of the \( g_{\text{OO}}(r) \) function's first peak, which appears slightly more pronounced.\\
In Table~\ref{dist}, we include the density (\(d\), in g/cm\(^3\)) of the amorphous Na\(_{3}\)OCl models simulated by FPMD and GAP. We remind that FPMD carried in the NVT ensemble were performed at a equilibrated reduced density of about 7\%  relative to the crystalline phase (corresponding to a volume expansion of +5\%), consistent with our previous work~\cite{pham_2023b}.
This adjustment minimized the stress tensor to values closed to 0 GPa for an energy cutoff of 160 Ry for the plane wave expansion. To further evaluate the performance of the GAP potential concerning density, we conducted MD simulations in the NPT ensemble at 300 K and 1 atm, averaging the final models' volume over 100 ps. The resulting error was less than 1.5\%, further confirming the reliability of the developed GAP potential.\\

\begin{figure}[!hb]
\begin{center}
\centering\includegraphics[width=0.8\linewidth]{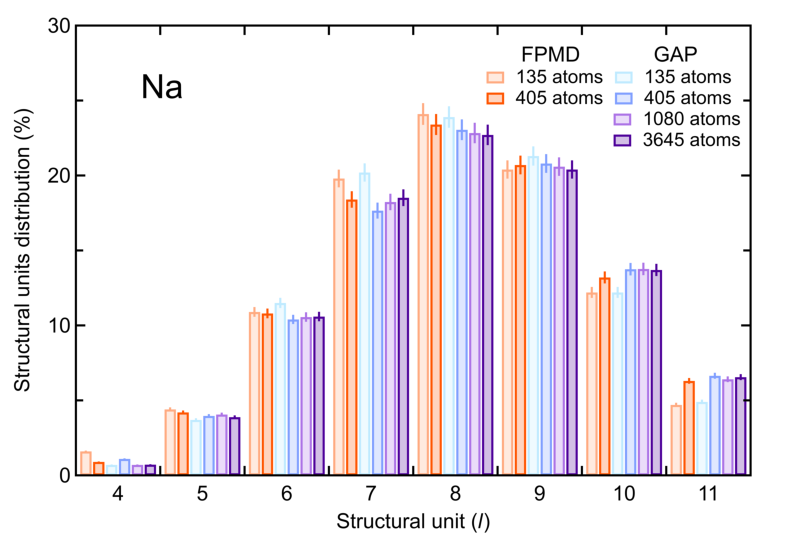}
\setstretch{2}
\caption{Distributions of the structural units for Na of amorphous
Na$_3$OCl as a function of the number of neighbors in each unit ($l$). Values obtained by using the definition of “total” coordination $n^t_{ij}$ , given in Section IVC.}
\label{SunitNa}
\end{center}
\end{figure}

\begin{table*}[!ht]
\centering
\setstretch{2}
\caption{Distribution of the individual cation-anion n$_\alpha(l)$ structural units where an atom of species $\alpha$ (Na, Cl or O) is $l$-fold coordinated to a counter ion computed for glassy Na$_3$OCl (see definition 1, Sec. ``Total and partial coordination numbers'').
In bold are reported the total percentages determined for each $l$-fold coordination. These quantities have been calculated including neighbours separated by a cutoff corresponding to the first minimum in the $g_{\alpha\beta}(r)$. For the present work, the individual pair cutoffs used are 3.30, 3.94, 4.10, 5.06, 5.51, and 5.39~{\AA} for, respectively, the Na-O, Na-Cl, Na-Na, O-O, Cl-Cl and O-Cl distances. A total cutoff of 3.30~{\AA} was defined from the total \( g_{\alpha \beta}(r) \). Only fraction greater than 1.5~\% are reported. Error bars are given in parenthesis. }
\begin{tabular}{lrcc|cccc}
\hline
\hline
   &    & \multicolumn{6}{c}{Proportions \=n$_\alpha(l)$ (\%)}\\
   &  & \multicolumn{2}{c}{FPMD} & \multicolumn{4}{c}{GAP}\\
   \multicolumn{2}{c}{n. atoms}    & 135    & 405 & 135 & 405 & 1080 & 3645\\
\hline
\textbf{Na} & & & &&&&\\
$l$=3 &  & \textbf{15.1}(0.8)&\textbf{14.7}(0.8)  & \textbf{10.9}(1.5) &\textbf{14.4}(0.9)& \textbf{14.4}(0.6) &\textbf{14.3}(0.3)\\
 &  O$_2$Cl$_1$   & 10.5 & 7.3  & 6.9 &8.8&8.7&7.8\\
 &  O$_3$     & 3.9 & 6.3 &  3.4&4.5&4.6&5.3\\
 $l$=4 &  & \textbf{66.4}(3.2)&\textbf{62.1}(0.9)   &  \textbf{72.9}(2.6)&\textbf{67.8}(2.0)&\textbf{63.2}(1.1)& \textbf{59.90}(0.6)\\
 &  O$_2$Cl$_2$   & 35.9 & 28.2 & 36.5&34.2&31.4& 29.5\\
&  O$_3$Cl$_1$    & 15.1 &18.2& 19.9&16.9&16.7&14.4\\
 &  O$_1$Cl$_3$   & 10.2  & 10.3& 14.2&12.1&11.0&12.0\\
 &  O$_4$         & 4.3  & 4.7 & 2.1&3.9&4.0&3.9\\
 $l$=5 &  & \textbf{16.6}(2.8)& \textbf{19.7}(0.8) & \textbf{14.9}(1.8)&\textbf{17.0}(0.8)&\textbf{20.0}(0.7)&  \textbf{17.5}(0.4)\\
 &  O$_2$Cl$_3$   & 6.5 &7.1  &  7.0&6.7&8.5&8.3\\
&  O$_1$Cl$_4$    & 5.3 & 7.9 &  5.1&5.7&6.7&5.1\\
 &  O$_3$Cl$_2$   & 1.7  & 3.4 &  1.8&2.8&3.4&3.0\\
  $l$=6 &  & \textbf{2.1}(1.2)& \textbf{2.1}(0.2) &  \textbf{1.1}(0.4)&\textbf{0.6}(0.2)&\textbf{2.2}(0.3)&\textbf{1.7}(0.3) \\
\hline
\textbf{O} & & & &&&&\\
$l$=5 & Na$_5$ & \textbf{1.2}(0.9)& \textbf{0.7}(0.5) &  \textbf{1.1}(0.8)&\textbf{1.0}(0.9)& \textbf{1.0}(0.5)& \textbf{1.1}(0.3)\\
$l$=6 & Na$_6$ & \textbf{66.8}(3.3)& \textbf{53.6}(0.9) & \textbf{71.3}(1.3)& \textbf{67.7}(2.1)& \textbf{61.5}(1.8)&  \textbf{62.2}(0.7)\\
$l$=7 & Na$_7$ & \textbf{30.0}(3.3)&\textbf{43.21}(1.1) &  \textbf{24.8}(1.9)&\textbf{27.7}(2.4)& \textbf{34.2}(1.9)&\textbf{32.6}(1.2) \\
$l$=8 & Na$_8$ & \textbf{1.5}(1)&\textbf{2.4}(1.4) &  \textbf{2.8}(1.1)&\textbf{2.6}(1.4)& \textbf{3.3}(1.0)&\textbf{4.0}(0.6) \\
\hline
\textbf{Cl} & & & &&&&\\
$l$=4 & Na$_4$ & \textbf{7.4}(1.4)&  \textbf{3.7}(0.9)&   \textbf{3.9}(1.9)&\textbf{6.2}(1.8)&\textbf{6.1}(1.3)&\textbf{5.9}(0.8)\\
$l$=5 & Na$_5$ & \textbf{31.0}(1.6)& \textbf{30.5}(1.0) &  \textbf{28.5}(2.3)&\textbf{31.9}(2.2)&\textbf{27.8}(1.8)&\textbf{28.0}(1.4) \\
$l$=6 & Na$_6$ & \textbf{41.7}(2.6)&\textbf{38.5}(0.9) &  \textbf{45.2}(1.9)&\textbf{42.3}(2.3)&\textbf{41.0}(2.1)& \textbf{41.4}(1.7)\\
$l$=7 & Na$_7$ & \textbf{16.9}(1.2)& \textbf{19.7}(0.9)&  \textbf{19.4}(1.8) &\textbf {17.3}(1.9)&\textbf{20.4}(1.5)& \textbf{20.0}(1.1)\\
$l$=8 & Na$_8$ & \textbf{2.0}(0.9)&\textbf{3.9}(0.5)  &  \textbf{2.6}(1.6)&\textbf{2.0}(1.4)&\textbf{4.1}(1.1)& \textbf{3.9}(0.6)\\
\hline   
\hline
\end{tabular}
\label{table_env2}
\end{table*}

\subsection{Total and partial coordination numbers}\label{totparcoo}

Further understanding of the Na\(_{3}\)OCl network is garnered through the analysis of total and partial coordination numbers (\(n_{i}\) and \(n_{ij}\), respectively), derived from the pair correlation functions. Specifically, \(n_{ij}\) is calculated by integrating the first peak of \(g_{ij}(r)\) up to the cutoff distance defined by the first minimum. The total coordination numbers for Na, O, and Cl are computed as follows: \(n_{Na} = n_{NaO} + n_{NaCl} + n_{NaNa}\), \(n_{O} = n_{ONa} + n_{OCl} + n_{OO}\), and \(n_{Cl} = n_{ClNa} + n_{OCl} + n_{ClCl}\).
A clear minimum following the first peak of \(g_{ij}(r)\) allows for an unambiguous definition of \(n_{ij}\), as the running coordination number shows a distinct plateau (e.g., Na-O pair~\cite{pham_2023b}). However, in the absence of this feature (e.g., Na-Na pair~\cite{pham_2023b}), \(n_{ij}\) becomes more sensitive to the cutoff choice. To address this issue, we use individual pair cutoffs determined by the minima following the first peak of \(g_{ij}(r)\) and a "total" cutoff—averaged across three amorphous models at 3.3 Å—following the second peak of \(g_{\text{tot}}(r)\), as in previous work~\cite{pham_2023b}. This approach considers all counter-ions (O and Cl) and some Na neighbors within Na's first coordination shell, facilitating a comparison with crystalline Na\(_{3}\)OCl coordination numbers.
Crystalline Na\(_{3}\)OCl is marked by Na atoms coordinated by six counter-ions (two O and four Cl) within the nearest-neighbor shell, with O atoms coordinated by six Na atoms and Cl atoms by twelve Na atoms. With our chosen cutoff (3.3 Å), each Na atom also neighbors eight Na atoms, leading to a "total" Na coordination of 14~\cite{Hippler1990,Pham2018}. This crystalline Na environment comprises 14 neighboring atoms, linked via Na-O-Na connections~\cite{pham_2023b}.
Accordingly, we analyzed amorphous Na\(_{3}\)OCl models' coordination numbers \(n_{ij}\) using "partial" \(n^p_{ij}\) (for cation-anion interactions, definition 2) and "total" \(n^t_{ij}\) coordination numbers (for a comprehensive neighbor overview, definition 1). This distinction helps characterizing each atom's surrounding structural units, revealing the network organization and providing insights into the specific chemical bonding nature going beyond average behaviors.
We present the calculated total coordination numbers (\(n^t_{\text{ij}}\)) for the three atomic species and the partial coordination numbers (\(n^p_{\text{ij}}\)) for Na in Table~\ref{dist}.\\
Focusing on the structural units, Fig.~\ref{SunitNa} illustrates their distribution based on the number of neighbors (\(l\)) for Na, utilizing definition 2. For Na atoms within a 3.3 Å cutoff, amorphous Na\(_{3}\)OCl shows a significantly different behavior compared to its crystalline counterpart. Specifically, the distribution of Na atom units spans a wide range (from 4 to 11) and is centered around \(l = 8.1\)-\(8.2\), with the count of neighboring counter-ions and Na neighbors diminishing from 6 and 8 to about 4.0-4.1, respectively. The FPMD and GAP models exhibit highly consistent results across all model sizes, within an averaged uncertainty bar of ±0.1 for coordination numbers. This consistency is indicative of a minimal size effect on this specific property.\\
By using definition 1, we report the distributions of structural units for Na, O, and Cl, focusing on interactions with what we designated as counter-ions. These distributions are given along with the detailed chemical composition breakdown of each structural unit for a specific \(l\).
\\
\subsection{Chemical identification of structural units}\label{chemiden}
The coordination units identified in the various FPMD and GAP models of amorphous Na\(_{3}\)OCl are detailed in Table~\ref{table_env2}. At the FPMD level, both the 135 and 405 atoms models exhibit similar trends, suggesting a predominantly fourfold first coordination shell for Na (\(\sim\)66\% and 62\%, respectively) with respect to cation-anion (counterion) interaction. The breakdown by chemical species for the 135 atoms model aligns closely with our prior findings~\cite{pham_2023b}, showcasing, for decreasing percentages, the units O\(_2\)Cl\(_2\) (\(\sim\)35.9\%), O\(_3\)Cl\(_1\) (\(\sim\)15.1\%), O\(_1\)Cl\(_3\) (\(\sim\)10.2\%), and O\(_4\) (\(\sim\)4.3\%). The newer 405 atoms FPMD model has a slightly different distribution, with O\(_2\)Cl\(_2\) at \(\sim\)28.2\% and O\(_3\)Cl\(_1\) at \(\sim\)18.2\%, while O\(_1\)Cl\(_3\) and O\(_4\) figures align with previous observations. Threefold and fivefold units amounts to about \(\sim\)15\% and \(\sim\)17-20\% respectively, with a notable occurrence of structures having two or one oxygen atom (O\(_2\)Cl\(_1\), O\(_2\)Cl\(_3\), and O\(_1\)Cl\(_4\)). The results from the 135 and 405 atoms models are highly consistent within the statistical error, indicating reduced variability for the larger model compared to the smaller one.
Our findings reveal that the structure of amorphous Na\(_{3}\)OCl markedly deviates from the molecular dynamics model proposed for glassy Li\(_{3}\)OCl, which suggests nanophase segregation into regions predominantly composed of Li\(_{2}\)O and LiCl~\cite{Hee19}. Specifically, when examining Na atoms that are fourfold or threefold coordinated in amorphous Na\(_{3}\)OCl, only approximately 8-11\%, in total, are solely coordinated with O atoms (O\(_3\) and O\(_4\)), and less than 1.5\% are exclusively coordinated with Cl atoms, fully in line with our previous finding~\cite{pham_2023b}.\\
For oxygen and chlorine, we focus solely on the analysis based on the total coordination number (\(n^{t}_{\text{O/Cl}}\)). Consistently with our previous findings~\cite{pham_2023b}, their local environments are predominantly made of counter-ions (Na), contrasting with the local environment of Na atoms.
The value of $n^{t}_{\rm O}$ in amorphous Na$_{3}$OCl is about $\sim$6.4-6.5 (Table~\ref{dist}), close to the one in crystalline Na$_{3}$OCl~\cite{Hippler1990,Pham2018,pham_2023}.
We observe that about 67\% of O atoms are coordinated by six Na atoms, and about 30\% by seven Na atoms in the 135 atoms FPMD model. There are notably different outcomes for the 405 atoms FPMD model, with about 54\% and 43\% of O atoms exhibiting six- and seven-fold coordination with Na, respectively (Table~\ref{table_env2}). Despite these variations in distribution units, the difference in the total coordination number for O (\(n^{t}_{\text{O}}\)) between the two model sizes is minor (6.4 vs. 6.5).
For Cl,  $n^{t}_{\rm Cl}$ is equal to $\sim$5.8 for both FPMD sizes, almost half of that found in the crystalline phase (12). 
Several structural units connected to Cl are noticeable in Table~\ref{table_env2}.  Indeed, approximately 31\% of Cl atoms in our simulations are coordinated by six Na atoms, while about 42-39\% are surrounded by five Na atoms, and roughly 17-20\% by seven Na atoms. The proportion of four- and eight-fold coordinated Cl atoms changes when comparing the 135 and 405 atoms models. Specifically, the 135 atoms model has about 7\% four-fold and 2\% eight-fold coordinated Cl atoms, whereas the 405 atoms model exhibits a lower presence of four-fold coordinated Cl (4\%) and a doubled content of eight-fold coordination (around 4\%).\\
For the GAP case, the 135 and 405 atoms models exhibit total coordination numbers for the three species (Na, O, and Cl) that are closely aligned with those derived from FPMD simulations, with values equal to 8.2, 6.3-6.5, and 5.8-5.9, respectively. However, a closer look to the individual structural unit distributions reveals that both GAP models feature distributions similar to those of the FPMD model for Na and Cl, within statistical uncertainty. In contrast, for O, the distributions in both GAP models more closely match the 135 atoms FPMD result. Within the GAP scheme, increasing the model size to 1080 and 3645 atoms results in a structural unit distributions of O atoms approaching values similar to those observed in the 405 atoms FPMD model.\\
Overall, the comprehensive analysis of amorphous Na$_3$OCl models through both FPMD and GAP methodologies has yielded significant insights into the structural characteristics of this material. The total and partial coordination numbers for Na, O, and Cl have revealed the subleties of the local atomic environments in amorphous Na$_3$OCl. The predominant coordination patterns for Na indicate a mix of threefold, fourfold, and higher coordination states, reflecting a complex network structure that diverges from the simpler models conjectured for glassy Li$_3$OCl. Both FPMD and GAP schemes exhibit remarkable consistency in the calculated total coordination numbers for all atomic species across different model sizes, confirming the occurrence pf minimal size effects. The distributions of structural units further emphasize the accuracy of GAP models in reproducing FPMD-derived structures, underscoring the potential of GAP as an efficient and reliable tool for studying amorphous antiperovskite materials. The analysis focused on O and Cl reveals that their local environments are predominantly made up of counter-ions (Na), a finding substantiating our previous study. This consistency across different models and sizes for O and Cl reveal the structural stability and uniformity of the amorphous Na$_3$OCl network. The detailed characterization of amorphous Na$_3$OCl provides valuable insights into its potential applications and stability. Understanding the coordination environments and the distribution of structural units is crucial for predicting material properties, such as ionic conductivity and stability, which are essential for applications in energy storage and conversion technologies and which will be focus of future effort.

\section{Conclusions}
The detailed study of amorphous Na$_3$OCl through FPMD and MLIP simulations has provided profound insights into its atomic structure and coordination environment. Our findings reveal a complex network structure with varying coordination patterns, significantly differing from the simpler conjectured models for glassy Li$_3$OCl. The consistency in total coordination numbers across different model sizes and methodologies highlights a negligible size effect on the material's structural properties. The GAP model, in particular, demonstrates remarkable fidelity in capturing the structural features of the amorphous Na$_3$OCl, as assessed by FPMD simulations, across an extended range of model sizes up to 3645 atoms. The analysis of oxygen and chlorine environments underscores their predominantly counter-ion composed local environments, differing from Na's local environment. This study not only advances our understanding of amorphous Na$_3$OCl's structural characteristics but also showcases the effectiveness of combining FPMD and MLIP approaches for electrolyte materials' analysis. Our findings lay a groundwork for future studies into the properties of amorphous materials and their potential technological applications, particularly in the field of energy storage and conversion technologies.

\section*{DATA AVAILABILITY}
The GAP potential, is freely available at \textcolor{red}{YYY} for non-commercial research. 
Representative configurations of Na$_3$OCl FPMD and MLIP models and raw trajectory data are available at the European Center of Excellence Novel Materials Discovery (CoE-NOMAD) repository (see ref.~\cite{nomad}).

\begin{acknowledgments}
This work was supported by the National Research Foundation
of Korea (NRF) grants funded by the Ministry of Science and ICT
(MSIT), government of South Korea (grant no.
2021R1F1A1048555). 
We acknowledge financial support from the Agence Nationale de la Recherche (ANR) within the framework of the project AMSES (ANR-20-CE08-0021) and the EUR-QMAT-ANR under the contract ANR-11-LABX-0058-NIE and ANR-17-EURE-0024 within the Investissement d’Avenir program ANR-10-IDEX-0002-02. We also acknowledge support from the Seed Money program (project MEDIA) of Eucor-The European
Campus. This project has also received funding from
the European Union’s Horizon 2020 research and innovation
programme under the Marie Skłodowska-Curie grant agreement number 847471 (QUSTEC). Calculations were performed
by using resources from GENCI (Grand Equipement National de
Calcul Intensif, grant no. A0120807670, A0140807670, and
A0140912441) and the Pole HPC Equipe@Meso of the Université de Strasbourg.
\end{acknowledgments}

{}

\begin{thebibliography}{}
\bibitem{zhao20}Zhao, Q., Stalin, S., Zhao, C. \& Archer, L. Designing solid-state electrolytes for safe, energy-dense batteries. {\em Nat. Rev. Mater.} \textbf{5} pp. 229-252 (2020)
\bibitem{fam19}Famprikis, T., Canepa, P., Dawson, J., Islam, M. \& Masquelier, C. Fundamentals of inorganic solid-state electrolytes for batteries. {\em Nat. Mater.}. \textbf{18}, 1278-1291 (2019)
\bibitem{Dawson2021}Dawson, J., Famprikis, T. \& Johnston, K. Antiperovskites for solid-state batteries: recent developments, current challenges and future prospects. {\em J. Mater. Chem.}. \textbf{9} pp. 18746-18772 (2021)
\bibitem{Wei22}Xia, W., Yang, Z., Feipeng, Z., Keegan, A., Ruo, Z., Shuai, L., Ruqiang, Z., Yusheng, Z. \& Xueliang, S. Antiperovskite electrolytes for solid-state batteries. {\em Chem. Rev.}. \textbf{122} pp. 3763-3819 (2022)
\bibitem{Shao2019}Shao, T., Liu, C., Deng, W., Li, C., Wang, X., Xue, M. \& Li, R. Recent research on strategies to improve ion conduction in alkali metal‐ion batteries. {\em Batteries \& Supercaps}. \textbf{2}, 403-427 (2019)
\bibitem{Kim2022}Kim, K., Li, Y., Tsai, P., Wang, F., Son, S., Chiang, Y. \& Siegel, D. Exploring the synthesis of alkali metal antiperovskites. {\em Chem. Mater.}. \textbf{34}, 947-958 (2022)
\bibitem{Yang2021}Yang, K., Liu, D., Qian, Z., Jiang, D. \& Wang, R. Computational auxiliary for the progress of sodium-ion solid-state electrolytes. {\em ACS Nano}. \textbf{15}, 17232-17246 (2021)
\bibitem{Dawson2022}Goldmann, B., Clarke, M., Dawson, J. \& Islam, M. Atomic-scale investigation of cation doping and defect clustering in the antiperovskite Na$_3$OCl sodium-ion conductor. {\em J. Mater. Chem. A}. \textbf{10} pp. 2249-2255 (2022)
\bibitem{Fan17}Fang, H. \& Jena, P. Li-rich antiperovskite superionic conductors based on cluster ions. {\em Proc. Natl. Acad. Sci. U.S.A }. \textbf{114} pp. 11046-11051 (2017)
\bibitem{Den21}Deng, Z., Ni, D., Chen, D., Bian, Y., Li, S., Wang, Z. \& Zhao, Y. Antiperovskite materials for energy storage batteries. {\em InfoMat}. \textbf{4} pp. 2567-3165 (2022)
\bibitem{Wang2015}Yonggang, W., Wang, Q., Liu, Z., Zhou, Z., Li, S., Zhu, J., Zou, R., Wang, Y., Lin, J. \& Zhao, Y. Structural manipulation approaches towards enhanced sodium ionic conductivity in Na-rich antiperovskites. {\em J. Power Sources}. \textbf{293} pp. 735-740 (2015)
\bibitem{Oh2021}Oh, J., He, L., Chua, B., Zeng, K. \& Lu, L. Inorganic sodium solid-state electrolyte and interface with sodium metal for room-temperature metal solid-state batteries. {\em Energy Storage Mater.}. \textbf{34} pp. 28-44 (2021)
\bibitem{Dawson2018}Dawson, J., Chen, H. \& Islam, M. Composition screening of lithium-and sodium-rich antiperovskites for fast-conducting solid electrolytes. {\em J. Phys. Chem. C}. \textbf{122}, 23978-23984 (2020)
\bibitem{Ahiavi2020}Ahiavi, E., Dawson, J., Kudu, U., Courty, M., Islam, M., Clemens, O., Masquelier, C. \& Famprikis, T. Mechanochemical synthesis and ion transport properties of Na$_3$OX (X= Cl, Br, I and BH$_4$) antiperovskite solid electrolytes. {\em J. Power Sources}. \textbf{471} pp. 228489 (2020)
\bibitem{Xu2020}Xu, X., Hui, K., Hui, K., Wang, H. \& Liu, J. Recent advances in the interface design of solid-state electrolytes for solid-state energy storage devices. {\em Mater. Horizons}. \textbf{7}, 1246-1278 (2020)
\bibitem{Zheng2021}Zheng, J., Perry, B. \& Wu, Y. Antiperovskite superionic conductors: A critical review. {\em ACS Mater. Au}. \textbf{1}, 92-106 (2021,11), 
\bibitem{Bra17}Braga, M., Grundish, N., Murchison, A. \& Goodenough, J. Alternative strategy for a safe rechargeable battery. {\em Energy \& Environ. Sci.}. \textbf{10}, 331-336 (2017,1)
\bibitem{Bra14}Braga, M., Ferreira, J., Stockhausen, V., Oliveira, J. \& El-Azab, A. Novel Li$_3$OCl based glasses with superionic properties for lithium batteries. {\em J. Mater. Chem. A}. \textbf{2}, 5470-5480 (2014,3), 
\bibitem{Bra16a}Braga, M., Murchison, A., Ferreira, J., Singh, P. \& Goodenough, J. Glass-amorphous alkali-ion solid electrolytes and their performance in symmetrical cells. {\em Energy \& Environ. Sci.}. \textbf{9}, 948-954 (2016)
\bibitem{Bra16b}Braga, M., Ferreira, J., Murchison, A. \& Goodenough, J. Electric dipoles and ionic conductivity in a Na$^+$ glass electrolyte. {\em J. Electroch. Soc.}. \textbf{164}, A207-A213 (2017)
\bibitem{Tia18}Tian, Y., Ding, F., Zhong, H., Liu, C., He, Y., Liu, J., Liu, X. \& Xu, Q. Li$_{6.75}$La$_{3}$Zr$_{1.75}$Ta$_{0.25}$O$_{12}$@amorphous Li$_3$OCl composite electrolyte for solid state lithium-metal batteries. {\em Energy Storage Mater.}. \textbf{14} pp. 49-57 (2018)
\bibitem{Tia2019}Tian, Y., Ding, F., Sang, L., He, Y., Liu, X. \& Xu, Q. Excellent lithium metal anode performance via in-situ interfacial layer induced by Li$_{6.75}$La$_{3}$Zr$_{1.75}$Ta$_{0.25}$O$_{12}$@ Amorphous Li$_3$OCl composite solid electrolyte. {\em Int. J. Electrochem. Sci}. \textbf{14} pp. 4781-4798 (2019)
\bibitem{Gao2021}Gao, Y., Sun, S., Zhang, X., Liu, Y., Hu, J., Huang, Z., Gao, M. \& Pan, H. Amorphous dual‐layer coating: enabling high li‐ion conductivity of non‐sintered garnet‐type solid electrolyte. {\em Adv. Funct. Mater.}. \textbf{31}, 2009692 (2021)
\bibitem{Tang22}Zhenghuan, T., Choi, J., Lopez, L., Co, A., Christopher Brooks, Jay Sayre \& Kim, J. Optimization of the Li$_3$BO$_3$ glass interlayer for garnet-based all-solid-state lithium–metal batteries. {\em ACS App. Energy Mater.}. \textbf{5} pp. 12132-12142 (2022)
\bibitem{Han18}Hanghofer, I., Redhammer, G., Rohde, S., Hanzu, I., Senyshyn, A., Wilkening, H. \& Rettenwander, D. Untangling the structure and dynamics of lithium-rich antiperovskites envisaged as solid electrolytes for batteries. {\em Chem. Mater.}. \textbf{30}, 8134-8144 (2018,11)
\bibitem{pham_2023}Pham, T., Choi, W., Shafique, A., Kim, H., Shim, M., Min, K., Son, W., Jang, I., Kim, D., Boero, M., Massobrio, C., Ori, G., Lee, H. \& Shin, Y. Structural-stability study of antiperovskite Na$_3$OCl for Na-rich solid electrolyte. {\em Phys. Rev. App.}. \textbf{19}, 034004 (2023)
\bibitem{Ste18}Steingart, D. \& Viswanathan, V. Comment on “Alternative strategy for a safe rechargeable battery” by M. H. Braga, N. S. Grundish, A. J. Murchison and J. B. Goodenough, Energy Environ. Sci., 2017, 10, 331–336. {\em Energy \& Environ. Science}. \textbf{11}, 221-222 (2018)
\bibitem{Fri23}Frith, J., Lacey, M. \& Ulissi, U. A non-academic perspective on the future of lithium-based batteries. {\em Nat. Commun.}. \textbf{14}, 420 (2023)
\bibitem{pol22}Poletayev, A., Dawson, J., Islam, M. \& Lindenberg, A. Defect-driven anomalous transport in fast-ion conducting solid electrolytes. {\em Nat. Mater.}. \textbf{21}, 1066-1073 (2022)
\bibitem{pham_2023b}Pham, T., Guerboub, M., Bouzid, A., Boero, M., Massobrio, C., Shin, Y.H., Ori, G. Unveiling the structure and ion dynamics of amorphous Na$_{3-x}$OH$_x$Cl antiperovskite electrolytes by first-principles molecular dynamics. {\em J. Mater. Chem. A}. \textbf{11}, 22922-22940 (2023)
\bibitem{Gao2023}Gao, L., Jiangyang, P., Longbang, D., Jinlong, Z., Liping, W., Song, G., Ruqiang, Z., Le, K., Songbai, H. \& Zhao, Y. Neutron diffraction for revealing the structures and ionic transport mechanisms of antiperovskite solid electrolytes.. {\em Chin. J. Chem.}. \textbf{42} pp. 100048 (2023)
\bibitem{Shah21}Shah, A., Shutt, R., Smith, K., Hack, J., Neville, T., Headen, T., Brett, D., Howard, C., Miller, T. \& Cullen, P. Neutron studies of Na-ion battery materials. {\em J. Phys. Mater.}. \textbf{4}, 042008 (2021,9
\bibitem{Tur23}Turanyi, R. \& Mukhopadhyay, S. Neutron spectroscopy and computational methods in investigation of Na ion battery materials: A perspective. arXiv:2304.08621. (2023)
\bibitem{cp85}Car, R. \& Parrinello, M. Unified approach for molecular dynamics and density-functional theory. {\em Phys. Rev. Lett.}. \textbf{55}, 2471-2474 (1985)
\bibitem{pbe}Perdew, J., Burke, K. \& Ernzerhof, M. Generalized gradient approximation made simple. {\em Phys. Rev. Lett.}. \textbf{77} pp. 3865 (1996)
\bibitem{TM91}Troullier, N. \& Martins, J. Efficient pseudopotentials for plane-wave calculations. {\em Phys. Rev. B}. \textbf{43}, 1993-2006 (1991)
\bibitem{Nose1}Nosé, S. A molecular dynamics method for simulations in the canonical ensemble. {\em Mol. Phys.}. \textbf{52}, 255-268 (1984)
\bibitem{Nose2}Nosé, S. A unified formulation of the constant temperature molecular dynamics methods. {\em J. Chem. Phys.}. \textbf{81}, 511-519 (1984)
\bibitem{Nose3}Massobrio, C. and I., Amiehe Essomba and M., Boero and C., Diarra and M., Guerboub and K., Ishisone and A., Lambrecht et al. On the actual difference between the Nosé and the Nosé–Hoover thermostats: A critical review of canonical temperature control by molecular dynamics. {\em Phys. Status Sol. (b)}. \textbf{261}, 2300209 (2024)
\bibitem{Massobrio2022}
C.~Massobrio, Amorphous materials via atomic-scale modeling, in: The structure of amorphous materials using molecular dynamics, 2053-2563, IOP Publishing, pp. 2--1 to 2--27, (2022)
\bibitem{Tucker92}Martyna, G., Klein, M. \& Tuckerman, M. Nosé–Hoover chains: The canonical ensemble via continuous dynamics. {\em J. Chem. Phys.}. \textbf{97}, 2635-2643 (1992)
\bibitem{Parr92}Blochl, P. \& Parrinello, M. Adiabaticity in first-principles molecular dynamics. {\em Phys. Rev. B}. \textbf{45}, 9413-9416 (1992)
\bibitem{Hippler1990}Hippler, K., Sitta, S., Vogt, P. \& Sabrowsky, H. Structure of Na$_3O$Cl. {\em Acta Cryst.}. \textbf{463} pp. 736-738 (1990)
\bibitem{cpmd} CPMD, copyright 1990-2023 by IBM Corp. and 1994-2001 by Max Planck Institute, Stuttgart. https://github.com/CPMD-code, (2023)
\bibitem{Bar10}Bartók, A., Payne, M., Kondor, R. \& Csányi, G. Gaussian approximation potentials: The accuracy of quantum mechanics, without the electrons. {\em Phys. Rev. Lett.}. \textbf{104} (2010)
\bibitem{Bernstein2019}Bernstein, N., Bhattarai, B., Csányi, G., Drabold, D., Elliott, S. \& Deringer, V. Quantifying chemical structure and machine-learned atomic energies in amorphous and liquid silicon. {\em 	Angew. Chem.}. \textbf{131}, 7131-7135 (2019)
\bibitem{Bar14}Szlachta, W., Bartók, A. \& Csányi, G. Accuracy and transferability of Gaussian approximation potential models for tungsten. {\em Phys. Rev. B}. \textbf{90}, 104108 (2014)
\bibitem{Bar15}Bartók, A. \& Csányi, G. Gaussian approximation potentials: A brief tutorial introduction. {\em Int. J. Quantum Chem.}. \textbf{115}, 1051-1057 (2015)
\bibitem{del20}Delaizir, G., Piarristeguy, A., Pradel, A., Masson, O. \& Bouzid, A. Short range order and network connectivity in amorphous AsTe$_3$: a first principles, machine learning, and XRD study. {\em Phys. Chem. Chem. Phys.}. \textbf{22}, 24895-24906 (2020)
\bibitem{unruh22}Unruh, D., Meidanshahi, R., Goodnick, S., Csányi, G. \& Zimányi, G. Gaussian approximation potential for amorphous Si : H. {\em Phys. Rev. Mater.}. \textbf{6}, 065603 (2022)
\bibitem{Der21}Deringer, V., Bartók, A., Bernstein, N., Wilkins, D., Ceriotti, M. \& Csányi, G. Gaussian process regression for materials and molecules. {\em Chem. Rev.}. \textbf{121}, 10073-10141 (2021)
\bibitem{kla23}Klawohn, S., James P., D., James R., K., Gábor, C., Miguel \& Albert, P. Gaussian approximation potentials: Theory, software implementation and application examples. {\em J. Mater. Chem. A}. \textbf{159} pp. 12 (2023)
\bibitem{Bar13}Bartók, A., Kondor, R. \& Csányi, G. On representing chemical environments. {\em Phys. Rev. B}. \textbf{87} (2013)
\bibitem{liu20}Liu, Y., Yang, J., Xin, G., Liu, L., Csányi, G. \& Cao, B. Machine learning interatomic potential developed for molecular simulations on thermal properties of $\beta$-Ga$_2$O$_3$. {\em J. Chem. Phys.}. \textbf{153}, 144501 (2020)
\bibitem{prodan2005}Prodan, E. \& Kohn, W. Nearsightedness of electronic matter. {\em Proc. Natl. Acad. Sci. U.S.A}. \textbf{102}, 11635-11638 (2005)
\bibitem{siva20}Sivaraman Machine-learned interatomic potentials by active learning: amorphous and liquid hafnium dioxide.. {\em Npj Comput. Mater.}. \textbf{6} pp. 104 (2020)
\bibitem{der18}Deringer, V.Data-driven learning of total and local energies in elemental boron. {\em Phys. Rev. Lett.}. \textbf{120} pp. 156001 (2018)
\bibitem{bar18}Bartók, A. Machine learning a general-purpose interatomic potential for silicon. {\em  Phys. Rev. X}. \textbf{8} pp. 041048 (2018)
\bibitem{lammps_22}Thompson, A., Aktulga, H., Berger, R., Bolintineanu, D., Brown, W., Crozier, P., Veld, P., Kohlmeyer, A., Moore, S., Nguyen, T., Shan, R., Stevens, M., Tranchida, J., Trott, C. \& Plimpton, S. LAMMPS - a flexible simulation tool for particle-based materials modeling at the atomic, meso, and continuum scales. {\em Comput. Phys. Commun.}. \textbf{271} (2022)
\bibitem{wright1993}Wright, A. The comparison of molecular dynamics simulations with diffraction experiments. {\em J. Non-Cryst. Solids}. \textbf{159}, 264-268 (1993)
\bibitem{Pham2018}Pham, T., Samad, A., Kim, H. \& Shin, Y. Computational predictions of stable phase for antiperovskite Na$_3$OCl via tilting of Na$_6$O octahedra. {\em J. App. Phys.}. \textbf{124}, 164106 (2018)
\bibitem{Choi2022}Choi, Y., C. Moyses Araujo \& Lizarraga, R. Amorphisation-induced electrochemical stability of solid-electrolytes in Li-metal batteries: The case of Li$_3$ClO. {\em J. Power Sources}. \textbf{521} pp. 230916 (2022)
\bibitem{Hee19}Heenen, H., Voss, J., Scheurer, C., Reuter, K. \& Luntz, A. Multi-ion conduction in Li$_3$OCl glass electrolytes. {\em J. Phys. Chem. Lett.}. \textbf{10}, 2264-2269 (2019)
\bibitem{nomad}Data deposited and available at NOMAD repository. https://dx.doi.org/10.17172/NOMAD/202Y.03.06-1, (2024)

\end{thebibliography}
\end{document}


\doublespacing
\begin{Center}
\Centering {SUPPLEMENTARY MATERIAL}
\end{Center}
\vspace{1cm}

\title{Structural properties of amorphous Na$_3$OCl electrolyte by first-principles and machine learning molecular dynamics}

\author{Tan-Lien Pham}
\altaffiliation{Current address: Department of Chemistry and Biochemistry Florida State University Tallahassee, FL 32304, USA}
\affiliation{Multiscale Materials Modeling Laboratory, Department of Physics, University of Ulsan, Ulsan 44610, Republic of Korea}
\author{Mohammed Guerboub}
\affiliation{Universit\'e de Strasbourg, CNRS, Institut de Physique et Chimie des Mat\'eriaux de Strasbourg, UMR 7504, F-67034 Strasbourg, France}
\affiliation{ADYNMAT CNRS consortium, F-67034 Strasbourg, France}
\author{Steve Dave Wansi Wendj}
\affiliation{Universit\'e de Strasbourg, CNRS, Institut de Physique et Chimie des Mat\'eriaux de Strasbourg, UMR 7504, F-67034 Strasbourg, France}
\affiliation{ADYNMAT CNRS consortium, F-67034 Strasbourg, France}
\author{Assil Bouzid}
\affiliation{Institut de Recherche sur les C\'eramiques, UMR 7315 CNRS-Univesit\'e de Limoges, Centre Europ\'een de la C\'eramique, 12 rue Atlantis 87068 Limoges Cedex, France}
\author{Christine Tugene}
\affiliation{Universit\'e de Strasbourg, CNRS, Institut de Physique et Chimie des Mat\'eriaux de Strasbourg, UMR 7504, F-67034 Strasbourg, France}
\author{Mauro Boero}
\altaffiliation{Current address: CNRS, Laboratoire ICube, Université de Strasbourg, UMR 7357, F-67037 Strasbourg, France}
\affiliation{Universit\'e de Strasbourg, CNRS, Institut de Physique et Chimie des Mat\'eriaux de Strasbourg, UMR 7504, F-67034 Strasbourg, France}
\affiliation{ADYNMAT CNRS consortium, F-67034 Strasbourg, France}
\author{Carlo Massobrio}
\altaffiliation{Current address: CNRS, Laboratoire ICube, Université de Strasbourg, UMR 7357, F-67037 Strasbourg, France}
\affiliation{Universit\'e de Strasbourg, CNRS, Institut de Physique et Chimie des Mat\'eriaux de Strasbourg, UMR 7504, F-67034 Strasbourg, France}
\affiliation{ADYNMAT CNRS consortium, F-67034 Strasbourg, France}
\author{Young-Han Shin}
\affiliation{Multiscale Materials Modeling Laboratory, Department of Physics, University of Ulsan, Ulsan 44610, Republic of Korea}
\author{Guido Ori}\email[]{Author to whom correspondence should be addressed: guido.ori@cnrs.fr}
\affiliation{Universit\'e de Strasbourg, CNRS, Institut de Physique et Chimie des Mat\'eriaux de Strasbourg, UMR 7504, F-67034 Strasbourg, France}
\affiliation{ADYNMAT CNRS consortium, F-67034 Strasbourg, France}

\date{\today}

\maketitle

\newpage

\begin{table}[]
\caption{Hyperparameters of GAP MLIP.}
\begin{tabular}{lc}
\hline
\hline
\multicolumn{2}{c}{\textbf{GAP 2b Descriptor} }\\
Cutoff \AA    &5.0 \\
$\delta$      & 0.1   \\
Sparse method & uniform   \\
Covariance    & Gaussian \\
Sparse points &  50    \\
\hline
\multicolumn{2}{c}{\textbf{SOAP Descriptor}} \\
Cutoff \AA  & 5.5 \\
Cutoff width \AA& 1.0 \\
$\delta$    & 1.0  \\
Sparse method       & CUR  \\
Sparse points       & 800   \\
$l_max$             & 8.0 \\
$n_max$             & 10.0 \\
$\zeta$             & 4.0 \\
\hline
\multicolumn{2}{c}{\textbf{Number of descriptors} } \\
\textbf{2b descriptors}  &   \\
{   Na-Na} &309704 \\
{   O-Na}  &236648 \\
{   O-Cl}  &71582\\
{   O-O}   &36190\\
{   Na-Cl} &203074 \\
{   Cl-Cl} &36990 \\
{   Total} &   894188 \\
\textbf{SOAP descriptors} & \\
{   Na} & 16201\\
{   O}  & 5401 \\
{   Cl} & 5401 \\
{   Total} & 27003\\
\hline
\multicolumn{2}{c}{\textbf{Regularisation}} \\
$\sigma_{e}$ [eV]   & 0.008 \\
$\sigma_{f}$  [eV/\AA] & 0.05  \\
$\sigma_{v}$ [eV]  &  0.05  \\
\hline
\multicolumn{2}{c}{\textbf{Database Size\footnote{\centering The total number of configurations (400) is split  50:50 in the training and testing sets.}} (cells, atoms)} \\
1200 K   &100,13500 \\
800 K  &100,13500 \\
450 K  &100,13500\\
300 K  &100,13500\\
\hline
\multicolumn{2}{c}{\textbf{Errors} (TR, TE)\footnote{\centering TR: training; TE: testing.}}\\ %
Energies: MAE$_{e}$   & 0.93,1.59 meV/atom\\
Forces: MAE$_{f}$   & 0.04, 0.04 eV/\AA\\
Virials: MAE$_{v}$ & 2.44,3.64 meV/atom\\
\hline
\hline
\end{tabular}
\label{table_env}
\end{table}
